\documentclass[twocolumn,preprintnumbers,secnumarabic,amsmath,amssymb,superscriptaddress,nofootinbib]{revtex4}

\usepackage[mathscr]{eucal} 

\usepackage{graphicx}
\usepackage[caption=false]{subfig}
 \usepackage{dcolumn}
\usepackage{bm}
\def\bg{\begin{equation}}
\def\nd{\end{equation}}

\usepackage{slashed}

\newcommand{\be}{\begin{equation}} 
\newcommand{\ee}{\end{equation}} 
\newcommand{\bea}{\begin{eqnarray}} 
\newcommand{\eea}{\end{eqnarray}}

\begin{document}

\date{\today} 
 
\title{A UV complete model of Large $N$ Thermal QCD} 
 
\author{Fang Chen} \email[email: ]{fangchen@hep.physics.mcgill.ca}
\affiliation{Department of Physics, McGill University, 
Montr\'eal, QC, H3A 2T8, Canada} 
\author{Long Chen} \email[email: ]{long.chen2@mail.mcgill.ca}
\affiliation{Department of Physics, McGill University, 
Montr\'eal, QC, H3A 2T8, Canada} 
\author{Keshav Dasgupta}\email[email: ]{keshav@hep.physics.mcgill.ca}
\affiliation{Department of Physics, McGill University, 
Montr\'eal, QC, H3A 2T8, Canada} 
\author{Mohammed Mia} \email[email: ]{mm3994@columbia.edu}
\affiliation{Department of Physics, Columbia University, 
538 West 120th Street, New York, New York 10027, USA} 
\author{Olivier Trottier}\email[email: ]{olivier.trottier@mail.mcgill.ca}
\affiliation{Department of Physics, McGill University, 
Montr\'eal, QC, H3A 2T8, Canada}

 
\begin{abstract} 

Many recent works on large $N$ holographic QCD in the planar limit have not
considered UV completions, restricting exclusively towards analyzing the IR
physics. Due to this, the UV problems like Landau poles and divergences of
Wilson loops including instabilities at high temperatures have not been
addressed. In some of our recent papers, we have discussed a possible UV
completion, which is conformal in the UV and confining in the far IR, that
avoids the Landau poles and the Wilson loop divergences. In this paper we 
give a field theory realization of this including the complete RG flow. 
We extend our UV complete model to study scenarios both above and below the 
deconfinement temperature and argue how phase transition in our model should 
be understood. Interestingly, because of the UV completion, subtle issues 
like instability due to negative specific heat do not appear. We 
also briefly elucidate the advantages that our model may have over other 
models studying large $N$ thermal QCD.

\end{abstract} 
 
\maketitle 


\section{Introduction}

The gauge/gravity duality has so far proved to be a powerful technique to solve many strong coupling problems of large $N$ gauge theories, and especially large 
$N$ QCD, in the planar limit. The application of a {\it gravity} dual to understand strongly coupled gauge theory was, in retrospect, the 
next best thing to do. A simple way to see this would be to consider a particular gauge-theory defined on a $3+1$ dimensional 
{\it slice} at a certain energy scale $\Lambda$. 
Now imagine we stack up all the slices together, described at different energy scales, 
along an orthogonal direction (call it the ``radial'' direction $r$). This way we will get a {\it five} 
dimensional space that captures the full dynamics of a given gauge theory from the Ultra-Violet (UV), i.e large $r$, to the Infra-Red (IR), i.e small $r$. 
The ``radial'' direction would then 
obviously be the direction along which the energy would change, i.e the direction of the Renormalisation Group (RG) flow. For a Conformal Field Theory (CFT), the 
theory does not change along the radial direction\footnote{Assuming the usual behavior of the irrelevant operators.}
 and therefore could as well be defined at the boundary of the five-dimensional space. The scale invariance of the underlying gauge theory will restrict the geometry 
of the five-dimensional space to the Anti-deSitter (AdS) space \cite{maldacena}, 
although it would be interesting to argue that this is the unique choice\footnote{Furthermore, a Feynman diagram for any interaction between point-like particles, 
when {\it stacked up} as above, would look like an interaction between extended objects, i.e strings! This is basically the essence of using string (or gravity) duals to 
study gauge theories. It will be informative to make this more precise.}. 

However, for gauge theories with inherent RG flows, the situation will be different and it would be instructive to study the theories at various $r$ (although we could also 
restrict ourselves to the boundary again). 
The example that we are interested in is large $N$ QCD, which we expect 
to be asymptotically conformal\footnote{It is interesting that we demand conformal behavior in the UV and not asymptotic freedom. This is because 
the 'tHooft coupling $\lambda \equiv g^2_{YM} N$ approaches a constant in the limit $g^2_{YM} \to 0$ and $N \to \infty$. This way, 
the theory is actually asymptotically free in terms of $g^2_{YM}$ but conformal in terms of $\lambda$. Furthermore, we will demand $\lambda$ to be very large 
throughout the whole 
RG flow so that the gravity dual can be restricted to its classical limit.} 
in the UV and confining in the far IR. Specific geometries that do the jobs 
for both zero and non-zero temperatures were presented in
\cite{FEP, jpsi} although the details of the gauge theories were not presented there. 
In this paper we will fill up some of the gaps left in \cite{FEP, jpsi} and argue why we believe our choice of the gravity dual is better suited to 
study large $N$ thermal QCD (see also \cite{kirit} for another model that studies UV complete large $N$ thermal QCD from a bottom-up five-dimensional point of view). 

\section{The field theory from the gravity dual}

The gravity dual of a large $N$ thermal QCD above the deconfinement temperature, described using only a {\it flavored}
Klebanov-Strassler geometry \cite{KS} with a black-hole has
few ultra-violet (UV) problems. For example, there are Landau poles coming from the flavor branes, and the Wilson loops are generically 
UV divergent \cite{UVissues1}.
All these issues could be resolved if we properly augment the Klebanov-Strassler geometry, which we will henceforth call as the 
Ouyang-Klebanov-Strassler black-hole (OKS-BH) \cite{ouyang, FEP, jpsi}
geometry, with a suitable asymptotically Anti-de Sitter (AdS) space. As discussed in \cite{jpsi}, this augmentation can only be performed in the presence of an 
interpolating space and certain number of anti five-brane sources.    

The interpolating region, which we called region 2 in \cite{jpsi}, can be interpreted alternatively as the {\it deformation} 
of the neighboring geometry once we attach an AdS cap to the OKS-BH geometry. The OKS-BH geometry is in the range 
$r_h \le r \le r_{\rm min}$ (which we will call as region 1) and the AdS cap is the range $r > r_0$ (which we will call as region 3). 
Here $r_h$ is the horizon radius. The geometry in the 
range $r_{\rm min} \le r \le r_0$ is the deformation. Such deformations should be expected for all other UV caps 
advocated in \cite{FEP}. This construction was elaborated in some details in \cite{jpsi}. 
In this paper we will start with a gauge theory interpretation of 
background.

\begin{figure}[h]
		\label{Image002}
                \begin{center}
		\vspace{- 0.7 cm}
		\centering
		\subfloat[RG flow in the far IR]{\includegraphics[width=0.25\textwidth, angle = 90]{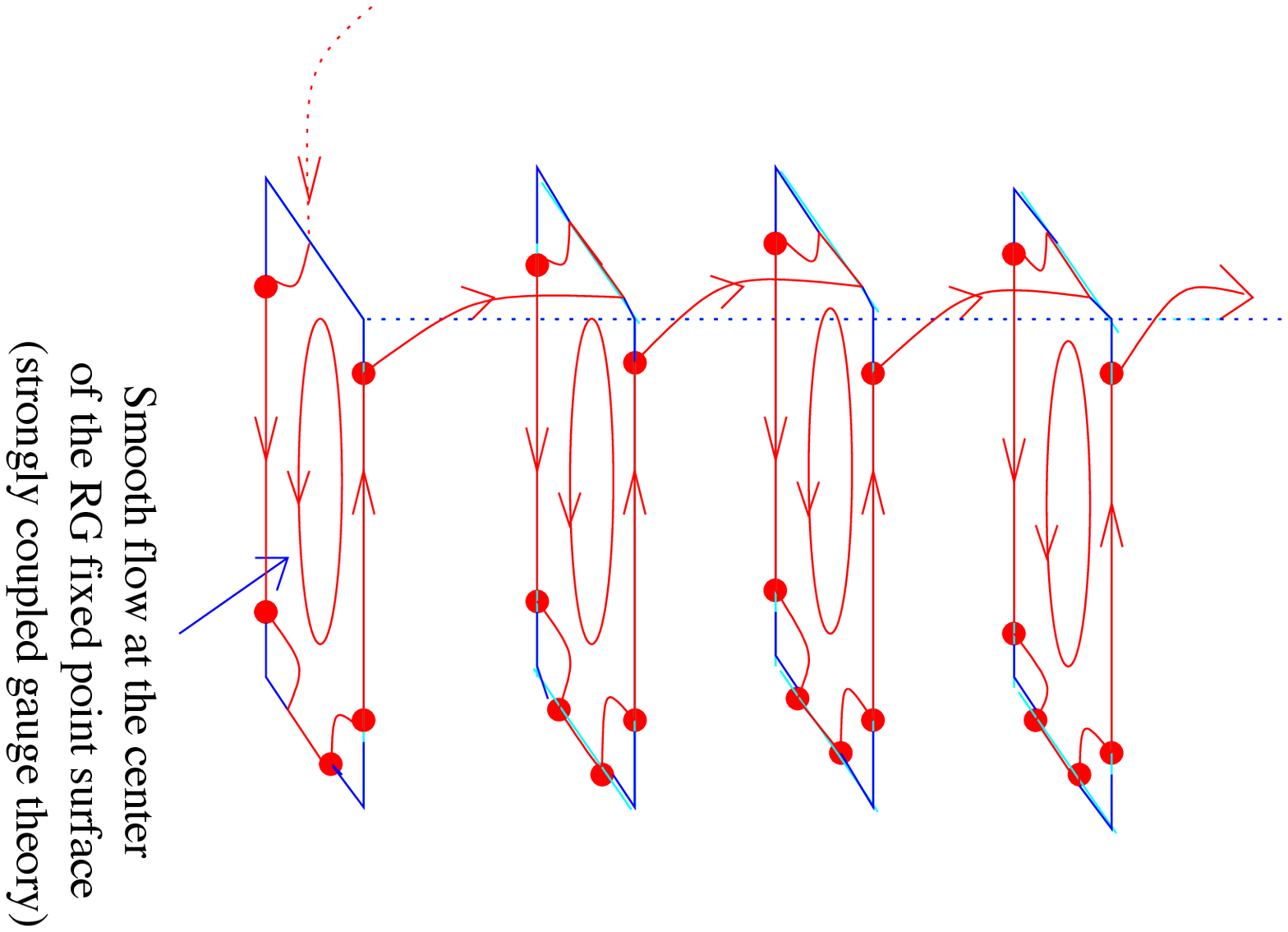}}
		\subfloat[RG flow in the far UV]{\includegraphics[width=0.15\textwidth]{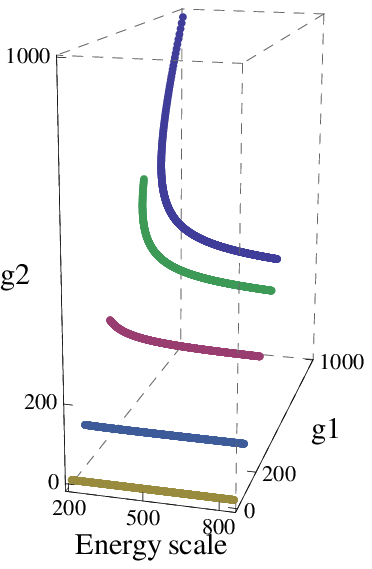}}
		\caption{{In the far IR, one may note that 
the cascading RG flow is never captured by the classical gravity theory. The classical supergravity description 
would capture only the smooth parts of the RG flow shown in (a) at the center of 
each slices. The vertical distances in (a) refer to the slices described using appropriate Seiberg dual descriptions. 
On the other hand, in (b) the RG flows all tend to go to zero at some UV 
scales. This is where 
the theory become conformal (all scales are chosen with $\alpha' = 1$). 
Note that, only the strongly coupled parts of figure (b) are captured by the classical supergravity description. }}
\end{center}
\end{figure}

For the UV region $r > r_0$ we expect the dual gauge theory to be $SU(N + M) \times SU(N + M)$ with fundamental 
flavors coming from the seven-branes. This is because addition of $M$
anti five-branes at the junction (i.e for $r > r_0$) with gauge fluxes on its world-volume tells us that the number of three-branes
degrees of freedom are $N + M$, where the $M$ and $N$ factors come from the presence of  $M$ five-branes anti-five-branes pairs and $N$ D3-branes. 
Furthermore, the $SU(N + M) \times
SU(N + M)$ gauge theory informs us that the gravity dual is approximately AdS, but has RG flows because of the 
fundamental flavors. In other words, the two couplings $g_1$ and $g_2$ of each gauge group would be approximately the same and exhibit
a walking RG flow.
At the scale $r = r_0$, we expect one of the gauge group 
to be Higgsed, so that we are left with $SU(N + M) \times SU(N)$. Now both gauge couplings flow at different rates 
and give rise to a cascade that is slowed down by the $N_f$ flavors. In the end, at far IR, we expect 
confinement at zero temperature.

The few calculations that we did in \cite{jpsi} regarding (a) the flow of $N$ and $M$ colors, 
(b) the RG flows, (c) the decay of the 
three-forms and (d) the behavior of the dual gravity background all support the gauge theory interpretation that we
gave above. What we haven't been able to demonstrate in \cite{jpsi, FEP} is the precise Higgsing that takes us to the cascading 
picture. From the gravity side, it is clear how this could be interpreted. From the gauge theory side, we will provide a brief derivation below.
But before we dwell on the details, let us see how the full Renormalisation Group (RG) flow would look like with the AdS cap.  

\subsection{Continuous RG Flow from UV to IR}

As mentioned above, the gravity dual should give us a
RG flow that allows us to see the UV conformal behavior and the IR confining behavior succinctly. However, there is a subtlety as shown in fig 1. The
cascading RG flow in the far IR, where the theory goes from one Seiberg fixed point to another, is in fact {\it not} seen in the dual gravity side because 
it runs between weakly coupled theories. Thus, what we see from the dual gravity side is a {\it smooth} RG 
flow\footnote{This also means that at any given scale $\Lambda$ there are in principle an infinite number of gauge theory descriptions available. Out of which, 
{\it one} of them 
might be the most useful description at that scale and is therefore captured by the classical supergravity analysis at $r = \Lambda$. 
For example, in the far IR, out of the many available gauge theory descriptions, it is the 
confining $SU(M)$ gauge theory (which is naturally strongly coupled) that is captured by the classical supergravity solution at small $r$. As a consequence, we 
expect that the definition of the number of colors at any given scale would become a little ambiguous.}, 
as depicted at the center of each 
slice in fig 1(a). 

The RG flow in the intermediate region, identified as region 2, is more involved and will be discussed in details in \cite{toappear}. However this RG flow connects smoothly
to the RG flow in the AdS cap, called as region 3. The flow in region 3 approaches conformality where both couplings run at an equal rate as shown in fig 1(b). 

\begin{figure}[htb]\label{1234}
		\begin{center}
\includegraphics[height=3cm]{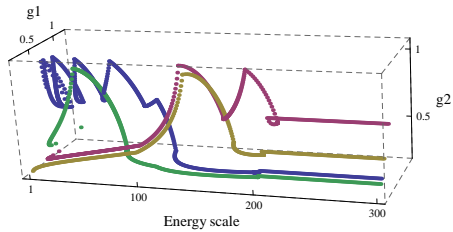}
		\caption{{A slightly unconventional way to represent the RG flow in our model. 
We get the complete RG flow by gluing the three regions altogether and using S-duality to transmute strong coupling into weak coupling. Starting from IR regime, once a particular coupling gets strong, a S-duality is performed to reverse the sign of the beta function associated with that coupling. This appears as the sharp edges in the figure above. From the UV region this can be seen in the following way:
The coupling starts as a constant in Region 3, when it gets to the transition point $r_0=200$, it has a small plateau region continuing in region 2, then it flows down to Region 1. The RG flow continues after the transition point $r_{min}=100$, but since the rate of change is fast more sharp corners appear in region 1. These are the points connected to their S-dual values. Eventually this reaches the smallest energy possible after which we expect linear confinement at low temperatures.
As before, all scales are chosen with $\alpha' \equiv 1$. }}
		\end{center}
		\end{figure}

The Beta functions are also easy to compute  to first order in $g_sN_f$
from the gravity dual. In Region 1 the two couplings at a scale $\Lambda$ run in the following way:
\bea\label{reg1rg}
&&\Lambda\frac{\partial}{\partial \Lambda}\left[\frac{4\pi}{{g_1}^2}+\frac{4\pi}{{g_2}^2}\right] = \frac{N_f}{8}
\left(\frac{6r^6+36a^2r^4}{r^6+9a^2r^4}\right)\Big\vert_{r=\Lambda}\\
&&\Lambda\frac{\partial}{\partial \Lambda}\left[\frac{4\pi}{{g_1}^2}-\frac{4\pi}{{g_2}^2}\right]=3M\left[1+\frac{3g_sN_f}{4\pi}\log(\Lambda^2+9a^2)\right]_{r=\Lambda}\nonumber
\eea
where the RHS of both equations is evaluated at  $r \equiv \Lambda$ in the gravity picture. The constant $a$ appearing above is the 
bare resolution parameter that one may set to zero\footnote{This is however not so above the deconfinement temperature. As shown recently in \cite{vaidya}, even if
we demand a vanishing bare resolution parameter, it'll get a contribution from the horizon radius $r_h$, such that $a \sim {\cal O}(r_h)$. Of course, on the gauge theory
side, the branes are still wrapped on vanishing cycle.}. 
In this limit, the RG flow is clearly the NSVZ RG flow \cite{nsvz}. On the other hand, 
in region 2, where we still have two couplings, the RG flow is highly non-trivial. This can be derived from the 
gravity dual where we see that the three-form fluxes play an important role in the running of the couplings \cite{toappear}:
\bea
\frac{8{\pi}^2}{{g_1}^2}=e^{-\Phi}\big[\pi-\frac{1}{2}+\frac{1}{2\pi}\int_{S^2} B_2\big]\\
\frac{8{\pi}^2}{{g_2}^2}=e^{-\Phi}\big[\pi+\frac{1}{2}-\frac{1}{2\pi}\int_{S^2} B_2\big]
\eea
Finally in region 3, the scenario is somewhat simpler. The two couplings flow approximately at the same rate 
and the flow is governed by the $N_f$ D7 and anti-D7 pairs that we keep in 
region 3 to cancel the Landau poles. These seven-branes are responsible for restoring the $SU(N_f) \times SU(N_f)$ chiral symmetry above the deconfinement
temperature (i.e when we insert a black-hole with a horizon radius $r_h$ \cite{FEP, jpsi}). The running of the coupling, which we call $g_{YM}$, is now:  
\bea
\Lambda\frac{\partial g_{\rm{YM}}}{\partial\Lambda} ={g_{\rm{YM}}}^3\sum_{n=1}^{\infty}\frac{\mathcal{D}_n}{\Lambda^{3n/2}}
\eea
where $
\mathcal{D}_n$ are all independent of $\Lambda$ and whose precise form will be derived in \cite{toappear}. 

\subsection {Higgsing}

In region 3, we have a $SU(N+M)\times SU(N+M)$ gauge group which breaks down
to $SU(N+M)\times SU(N)$ by the Higgs mechanism as we enter Region 1. We will study this mechanism in
two versions: supersymmetric and non-supersymmetric. Since the purpose is to
break the gauge group, we will ignore any fundamental matter fields in the following discussion.

Before moving ahead, let us see how we could justify the Higgs mechanism from the gravity perspective. The brane construction that reproduces the gauge theory should be 
understood on a 
scale-by-scale basis, so that the full RG flow could be reproduced in the gravity dual. Generically, we expect $N$ D3s and $M$ wrapped D5s on a vanishing two-cycle
of the conifold. Allowing a small resolution factor to the other two-cycle, we can distribute the anti-D5 branes on the resolved sphere such that they wrap the same vanishing 
two-cycle but are distributed on the other sphere. Similarly the D7 and anti-D7 branes are also 
distributed\footnote{The tachyons between D5 and anti D5-branes or between D7 and anti D7-branes can be cancelled by switching on appropriate gauge fluxes on the set of
anti branes. This 
phenomena is somewhat similar to the ones in \cite{bakkarch}. These gauge fluxes will create bound D3 and bound D5-branes respectively on the two set of brane anti-brane
systems. If oriented properly, 
the system would then be almost BPS in the zero-temperature case when the distance between the two set of branes is large (the multipole forces 
are heavily suppressed). To stabilize this completely, 
one may switch on three-form $H_{\rm NS}, H_{\rm RR}$ 
fluxes on the internal space (the axio-dilaton are already switched on). These $H$-fluxes
would not only change the moding of the strings between the branes but also stabilize the position of the branes, by generating perturbative and 
non-perturbative superpotential and giving masses to the 
scalar fields on the branes, along the lines of \cite{DRS, gorlich, bachas}. Alternatively for short distances, one 
may dissolve the anti-D5 branes in the D7 anti-D7 system in the way discussed in \cite{jpsi}, and then stabilize the seven-brane positions. In either case the physics 
would be the same.},   
over the resolved sphere via the Ouyang embedding 
\cite{jpsi}. 
This configuration is more intuitive
from the gravity dual side where the radial coordinate now becomes the scale of the theory. At a given scale we 
expect $M_\epsilon$ number of 
wrapped anti-D5 branes where $M_\epsilon = {Me^{\alpha(r-r_0)}\over 1 + e^{\alpha(r-r_0)}}$ with $r \sim {\cal O}(1/\epsilon)$ and $\alpha >> 1$. Then it is 
easy to see that the resulting gauge group becomes $SU(N+M_\epsilon) \times SU(N + M)$. Clearly, in the limit $\epsilon \to 0$, we recover the conformal gauge group. Therefore,
the anti-D5 branes appear to only affect one of the gauge groups in the product. In region 3, where $r >> r_0$, $M_\epsilon \approx M$, this tells us that the RG flow will be 
mostly due to the flavor seven-branes\footnote{The gauge group that actually appears in the far IR is $SU(N+M) \times SU(N) \times U(1)^M$, where the $U(1)$'s are from the
massless sector of the anti-D5 branes. However at low energies, once we integrate out the Higgs masses (i.e the strings between the D5 and the anti-D5 branes), 
these $U(1)$'s would be decoupled. Furthermore at strong coupling, where we 
expect the dual 
gravity description to hold, these $U(1)$'s will never appear. This in turn implies that the anti five-brane degrees of freedom should only be seen at high energies,
precisely in the way we predicted in \cite{jpsi}!}.   

The above construction then instructs us that the Higgsing process generating the cascade should simply be engineered 
by making some anti-D5 brane DOFs heavy, i.e by moving the anti-D5 branes
away from the $N$ D3 and the M wrapped D5 branes on the resolved sphere as we discussed above.
In a supersymmetric theory, where the UV completion is done by a ${\cal N} = 2$ theory, this process would mean moving the 
anti-D5 brane DOFs along the Coulomb branch, which in turn implies that the anti-D5 branes' world-volume scalar multiplets, transforming under a certain subgroup 
of $SU(N + M)$, will be responsible for the Higgsing mechanism. 

For the non-supersymmetric theory, this is rather easy to demonstrate. All we require is that 
the Higgs field $\phi$ should only transform under a certain subgroup of the first $SU(N + M)$ group. The Lagrangian is:
\begin{eqnarray}
\mathcal{L}=-\frac{1}{2}D^{\mu}\phi_k D_{\mu}\phi_k-V(\phi)-\frac{1}{4}F^{a\mu\nu}_iF^a_{i\mu\nu}
\end{eqnarray}
where $i = 1, 2$ refers to each $SU(N+M)$ copy in the product gauge group. $D_{\mu}\phi_k=\partial_{\mu}\phi_k-ig_1A^a_{1\mu}(T^a_1)_{kl}\phi_l$
with $g_1$, $A_{1\mu}$ and $T_1$ being the gauge coupling, gauge field and
generators of the first $SU(N+M)$ group respectively. We can choose the matrix representation of the $T_1$ generator properly to demand what subgroup of $SU(N+M)$ we want. 

Now we suppose the potential $V(\phi)$ is minimized at $\langle\phi_{i} \rangle \equiv v_i$. 
Then a generator $T_1^a$ is broken if $\left(T^a_1\right)_{ij}v_j\neq 0$. To see this, let's
write:
\begin{eqnarray}
\phi_i(x)=v_i+H_i(x)
\end{eqnarray}
where $H_i(x)$ is a real scalar field. The covariant derivative of $\phi_i$
is:
\begin{eqnarray}
D_{\mu}\phi_i = \partial_{\mu}H_i(x)-ig_1A^a_{1\mu}(T^a_1)_{ik}\left[v_k+H_k(x)\right]
\end{eqnarray}
and the kinetic term for $\phi_k$ becomes:
\begin{eqnarray}
&&-\frac{1}{2}D^{\mu}\phi_k D_{\mu}\phi_k =-\frac{1}{2}\partial^{\mu}H_k\partial_{\mu}H_k -\frac{1}{2}M_k^aM_k^bA^{a\mu}_1A^b_{1\mu}\nonumber\\
 && +M^a_kA^a_{1\mu}\partial^{\mu}H_k +ig_1A^a_{1\mu}H_i\left(T^a_1\right)_{ij}\partial^{\mu}H_j\nonumber\\ 
&&  +\frac{1}{2}g_1^2A^{a\mu}_1A^b_{1\mu}H_i\left(T_1^a\right)_{il}\left(T_1^b\right)_{lj}H_j\nonumber\\
&& + ig_1A^{a\mu}_1A^b_{1\mu}M_i^a\left(T^b_1\right)_{ij}H_j 
\end{eqnarray}
where $M_k^a=ig_1T^a_{kj}v_j$. It is obvious now that $A^a$ will get massive
if $T^a_{ij}v_j\neq0$ and thus the gauge group is broken. In our case, we
 only want to break $M$ of the generators. How this is done depends on the details of
 the potentials and the specific values of $N$ and $M$.

{}From the dual gravity, we expect to see $M$ anti-D5 branes at $r \to \infty$ so that the gauge theory is almost conformal. As the radial coordinate decreases, the number 
of anti-D5 branes become $M_\epsilon$, as given earlier. For $r << r_0$ we expect the number of anti-D5 branes to completely vanish so that the gauge group becomes 
$SU(N+M) \times SU(N)$ henceforth the cascading behavior begins. In figure 3 we have plotted the behavior of the function $f(r) = M_\epsilon/M$.   

\begin{figure}[htb]\label{ffunction}
		\begin{center}
\includegraphics[height=5cm,angle=-90]{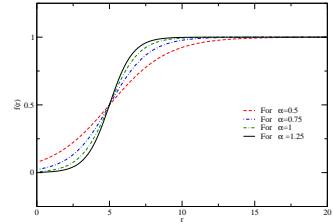}
		\caption{{A plot of the function $f(r) \equiv {e^{\alpha(r-r_0)}\over 1 + e^{\alpha(r-r_0)}}$ for $r_0 = 5$ in appropriate units, and for various 
choices of $\alpha$. For $r << r_0$, the function vanishes whereas it approaches unity for $r > r_0$.}}
		\end{center}
		\end{figure}

The supersymmetric case follows the same line of argument as above.
The general Lagrangian now is:
\begin{eqnarray}
\mathcal{L}=\int d^4\theta~ \mathcal{K}\bar{\Phi} e^V \Phi+\int d^2\theta \left[\mathcal{W}+\frac{1}{32\pi i}\tau {\rm tr}_fW^2_{\alpha}\right]+h.c.\nonumber\\
\end{eqnarray}
where $\left(\Phi, W_\alpha = -{1\over 4} \bar{D} \bar{D} D_\alpha V\right)$ 
are the appropriate ${\cal N} = 1$ chiral and the vector multiplets with ($\phi_k, A^a_{\mu}$) being the complex scalar and the vector fields in their
respective multiplets,
$\mathcal{K}$ is a gauge invariant K\"ahler potential, $\mathcal{W}$ is a gauge invariant superpotential, 
$\tau \equiv \frac{\vartheta}{2\pi}+i\frac{4\pi}{g^2}$ is the complexified gauge coupling with $\vartheta$-angle and ${\rm tr}_f$ 
is the trace in the fundamental representation. The FI terms don't appear because they are forbidden by the
non-Abelian gauge invariance. 

The scalar potential obtained by expanding the above Lagrangian in components is a sum of $F^2$ and $D^2$ terms. The $F$ and the $D$ terms  are:
\begin{eqnarray}
F_k = -\frac{\partial\bar{\mathcal{W}}}{\partial\phi_k}, ~~~~ D^a = \bar{\phi}_k\left(T^a_R\right)_{kl}\phi_l
\end{eqnarray}
where $T^a_R$ denotes the generator in the $R$ representation.
To preserve supersymmetry we must have $F_k =0$ and $D^a=0$. Actually in the absence
of FI terms whenever $F_k =0$ has a solution, $D^a=0$ always has a solution. So we
assume we already have a solution that satisfies $F_k=0$. This solution can break the
gauge symmetry as in the non-supersymmetric case. This can be seen in the following way.

Write down the relevant kinetic terms of the scalar component of the Higgs multiplet, as:
\begin{eqnarray}
\int d^4\theta ~\bar{\Phi}e^V\Phi=-\left\vert(\partial_{\mu}+igA^a_{\mu}T^a)\phi\right\vert^2+...
\end{eqnarray}
which is exactly the same as in the non-supersymmetric case. If the $F$-term solution leads to $M$ generators $T^a$ such 
that $T^a_{ij}v_j\neq 0$, then the gauge group is broken from
$SU(N+M)$ to $SU(N)$. How this happens again depends on the details of the $N$, $M$ values and the form of the superpotential.

\section{Phase transition and other applications}

Once we have the gauge theory description, it is time to extend our configuration to incorporate temperature. Two immediate scenarios present themselves: the confining 
theory at low temperatures and the theory above the deconfinement temperature. The process of going from one to another in the gravity dual will appear as the 
confinement to deconfinement phase transition in the large $N$ thermal QCD. 

In \cite{FEP, jpsi, vaidya}, the theory above the deconfinement temperature was studied in details. The high temperature phase was understood therein as the one coming from 
a black-hole with a horizon radius $r_h$ where the temperature was related to $r_h$. The scenario at low temperatures were not discussed in 
details in \cite{FEP, jpsi, vaidya}. Here, 
we will study these two phases 
and discuss their associate phase transition. More elaborations on this will be presented in \cite{fangmia}. 

Before actually computing the phase transition, let us discuss a couple of issues that may arise in equivalent scenarios dealing with large $N$ thermal QCD. The first  
issue is the stability at high temperatures. Stability is guaranteed by a positive specific heat $c_v$. A 
negative specific heat implies instability, which in fact turned out to be the case of many models that study large $N$ thermal QCD without a UV completion \cite{recent}.   
To assess the issue of stability, let us first define the specific heat in terms of the internal energy 
$E_{\rm int}$, the BH factor $g = 1-r_h^4/r^4$ and the temperatute $T$ in the following way:
\bea
c_v ~= ~ \left(\frac{\partial E_{\rm int}}{\partial T}\right)_V,\indent \mathrm{T} = \frac{g'}{4\pi\sqrt{h}}\Big|_{r_h}\simeq \frac{r_h}{\pi L^2}
\eea
where we have introduced the $AdS_5$ length scale $L$ in anticipation of the AdS cap, and 
the internal energy is given by the integral of the zeroth component of the stress tensor. 
To calculate the heat capacity, we have to know how much energy is encoded in the geometry. As $r\rightarrow\infty$, the space-time is 
approximately $AdS_5\times T^{1,1}$, where $T^{1,1}$ is the internal space. The internal energy of asymptotically $AdS_5$ space-time can be easily calculated using results 
from \cite{kostas1}. The total stress-energy tensor $T_{ij}$ is composed of stress-energy from the medium and 
the quarks. Quarks can be seen as excitations of the D7 branes. 
At the boundary, only the medium contributes to the stress-energy tensor\footnote{The contribution from $T_{00}$ and warp factor $h$ as $ r \rightarrow \infty$ is of zeroth 
order. When the AdS geometry is deformed, the stress-energy receives higher order corrections.}.
Thus, using the background above the deconfinement temperature given in \cite{FEP, jpsi, vaidya}, the internal energy, in terms of the string coupling $g_s$ and the 
Newton's constant $G_N$, becomes:
\bea
E_{\rm int} = \int d^3 x \sqrt{g}~T_{00}= \frac{\pi^2{r_h}^4 }{{g_s}^2 G_N}
\eea
which gives the following value for the specific heat:
\bea
c_v~=~ + \frac{4\pi^6 L^8}{{g_s}^2 G_N}\mathrm{T}^3
\eea
This means that the heat capacity is positive for positive temperatures, showing that the model is stable at high temperatures.

The second issue is slightly tangential to our interest but is nevertheless important enough that we clarify the scenario here. It was proposed recently in an 
interesting work \cite{manmort} that the confinement to deconfinement phase transition in the type IIA Sakai-Sugimoto model \cite{SS} {\it does not} proceed 
via the standard transition of a solitonic D4-brane to a black D4-brane, as proposed in \cite{Witten:1998zw, aharony}, but via a Gregory-Laflamme transition \cite{GL} 
from a solitonic D4-brane to a certain type IIB Euclideanized D3-brane configuration. In retrospect,
this conclusion may not be too surprising because the black D4-brane elegantly depicts
the {\it five}-dimensional deconfined phase but fails to do so
in the four-dimensional case once a certain energy scale is reached. Indeed in this model, there is no reason for integrating out the 
modes coming from the compact $S^1$ direction. Thus the Euclideanized D3-brane phase should be preferred at high temperatures. 
Unfortunately however, because the Mandal-Morita \cite{manmort} picture above the deconfined 
phase is {\it not} a configuration of black D3-branes, the usual computations of transfer coefficients, that rely on the dynamics of 
black-holes in these spaces, cannot be performed so easily.    

This is exactly where our model may have some distinct advantages. Since we are considering configurations of wrapped five-branes and anti five-branes on {\it vanishing}
two-cycle, the subtlety of Kaluza-Klein (KK) reduction will not
appear, and we should be able to go between solitonic D3 and black 
D3-branes. This
would then be the confinement to deconfinement phase transition for our case, which is of course the Hawking-Page \cite{Hawking:1982dh} 
transition. In the following we will first take a brief detour to explain the Sakai-Sugimoto limit of our model, before going into the discussion of phase transition in our 
set-up. More details will appear in \cite{toappear, fangmia}.

\subsection{The Sakai-Sugimoto limit}

{}From the above discussion, 
an interesting question at this stage would be to compare our type IIA dual picture with the Sakai-Sugimoto model\cite{SS}. For simplicity, let us
only consider the far IR picture where we have D5-branes wrapped on the vanishing two-cycle of the conifold. The vanishing cycle could be parametrised by ($\theta_1, \phi_1$)
and the other two-cycle is along ($\theta_2, \phi_2$). The $U(1)$ fibration of the conifold is $\psi$ and the radial direction is $r$. The D5-branes have a spacetime 
stretch along the usual $x^{0, 1, 2, 3}$ directions. T-dualising along the $\psi$ direction gives us D4-branes stretched between two NS5-branes along 
the $\psi$ circle \cite{DM}. Thus
the $\psi$ coordinate is like the $x^4$ coordinate of the Sakai-Sugimoto model. The difference now is that the D4-branes are 
stretched only along a fraction of the $\psi$ circle and between two
orthogonal NS5-branes. The D7 anti-D7-branes become D8 anti-D8-branes along ($x^{0, 1, 2, 3}, r$) and ${\bf P}_{\theta_1, \phi_1}^1 \times {\bf P}_{\theta_2, \phi_2}^1$ 
just like the Sakai-Sugimoto case, but 
with an instanton configuration on the two spheres that breaks the supersymmetry. Of course, one might now worry that, since we made $\psi$ non-contractible, the usual 
issue raised in \cite{manmort} should appear for our T-dual model too. However, note that the distance between the 
two NS5-branes could be made arbitrarily small\footnote{This depends on the choice of the $B_{\rm NS}$ field on the vanishing cycle \cite{DM}.} 
(without changing the size of the $\psi$ circle), so the issue raised in \cite{manmort} may appear only at very high 
temperatures! Thus even at arbitrarily high temperature, if we tune the distance between the two NS5-branes appropriately so that the $\psi$ modes are of very high energies, 
we might still be able to study the deconfined limit using the 
black D4-branes. Further details and explicit computations on this construction will be reported in \cite{toappear}.

\subsection{Phase transition}  

Phase transitions of $SU(N)$ gauge theory can be realized by spontaneous breaking of the center symmetry ${\bf Z}_N$. In the confined phase, ${\bf Z}_N$
symmetry is preserved and its associated order parameter, a temporal Wilson loop, is zero
(i.e $\langle W \rangle =0$). In the deconfined phase, ${\bf Z}_N$ symmetry is
spontaneously broken with $\langle W \rangle \neq 0$. In \cite{jpsi}, we computed $\langle W \rangle$ using the gravity description and showed that OKS-BH geometry with large
black holes give $\langle W \rangle \neq 0$ while the OKS geometry without black holes give $\langle W \rangle =0$. This indicates that extremal geometry is dual to confined
phase while non-extremal geometry corresponds to deconfined phase. 

Here we will obtain the critical temperature for confinement/deconfinement transition by computing the free enegy of extremal and non-extremal
geometries and identifying it with  the free energy of the gauge theory. We start with the on-shell type IIB supergravity action with
 appropriate
Gibbons-Hawking boundary terms and counter terms:  
\bea \label{action1}
{\cal S} =\beta E_{\rm free}= S_{IIB}+S_{GH}+ S_{\rm counter}
\eea
where $E_{\rm free}$ is the free energy,  
  $S_{\rm IIB}$ is the ten dimensional type IIB  Euclidean supergravity action including localized sources \cite{DRS}\cite{GKP}, $S_{GH}$ is the Gibbons-Hawking 
surface term \cite{Gibbons-Hawking} and $S_{\rm counter}$ is the counter term necessary to renormalize the action \cite{kostas1}\cite{FEP}\cite{fangmia}. 
Just like the case for AdS gravity discussed by Hawking and Page \cite{Hawking:1982dh} and subsequently by Witten \cite{Witten:1998zw}, the above action gives
 rise to both extremal
and non-extremal metric and both geometries can incorporate non-zero temperature of the dual gauge theory in the following way: Wick rotate $t\rightarrow
i\tau, \tau\in (0,\beta)$ and identify temperature $T$ as $T=1/\beta$. At a fixed temperature of the gauge theory, we
have two geometries $-$ extremal and non-extremal $-$  and the geometry with smaller on-shell action will be preferred. The free energy of the gauge
theory will then be given by the free energy of the geometry obtained through (\ref{action1}).  
Denoting the on-shell value of the action for the extremal geometry with ${\cal S}_1$ and the non-extremal geometry with
${\cal S}_2$, we compute the action difference in the absence of D7 branes and localized 
sources, i.e. $N_f=0$ and the axio-dilaton $\tau$ is a constant (i.e  without fundamental matter), as \cite{fangmia}:
\bea \label{actiondiff}
\triangle {\cal S} &=& {\cal S}_2- {\cal S}_1\nonumber\\
&=&\frac{g_sM^2\beta_2 V_8}{2\kappa_{10}^2N}\lim_{{\cal R}\rightarrow \infty}
  \left[\frac{r_h^4}{32}{\rm log}\left(\frac{{\cal R}}{r_h}\right)-\frac{5d r_h^4}{128}
  \right]\nonumber\\
\eea  
where $V_8$ is the volume of $R^3\times T^{1, 1}$, $T^{1, 1}$ being the base of the conifold with approximate radius $L=(g_sN)^{1/4}\sqrt{\alpha'}$, $N, M$ are number of D3
and D5 branes, ${\cal R}$ is the boundary value of $r$, and  $r_h$ is the black hole horizon radius. Here $d>0$ is a constant independent of
$N,M, g_s$ and depends on the boundary values of derivatives of the metric \cite{fangmia}. In obtaining (\ref{actiondiff}), we have only kept terms up to
linear order in $g_sM^2/N$ which is valid for $N\gg g_sM^2$ and the exact form of $S_{\rm counter}, S_{\rm GH}$  is presented in
\cite{fangmia}. The critical temperature is obtained by evaluating the critical horizon $r_h^c$ for which $\triangle S(r_h^c)=0$ and the
result is \cite{fangmia}: 
\bea \label{T_c}
r_h^c={\cal R} {\rm exp}\left(-\frac{5d}{4}\right), ~ T_c=\frac{1+{\cal O}\left(\frac{g_sM^2}{N}\right)}{\pi \;{\rm exp}\left(\frac{5d}{4}\right)
(g_sN)^{1/4}\sqrt{\alpha'}}\nonumber\\
\eea
where we have used the scaling ${\cal R}=L=(g_s N)^{1/4}\sqrt{\alpha'}\rightarrow \infty$.
For $T>T_c$, $\triangle S<0$, i.e the black hole geometry has lower free energy and thus preferred, while for $T<T_c$, $\triangle S>0$, i.e 
the extremal geometry is preferred. For
extremal geometry, one readily gets an entropy $s=-\frac{\partial E_{\rm free}}{\partial T}=0$, while for the black hole geometry:
\bea\label{bhento} 
s\sim N^2 T^3\left[1+ \frac{g_sM^2 b}{N} ~{\rm log}(LT)\right] 
\eea
at
lowest order in $g_sM^2/N$ and $b>0$ is a constant independent of $N,M,g_s$. Observe that when $M=0$, $\triangle {\cal S}=0, \forall r_h$ $-$ i.e extremal and
non-extremal action is equivalent for {\it all temperatures} of the boundary gauge theory. This is consistent with the field theory picture because 
the $M=0$ limit gives an $AdS_5\times T^{1, 1}$ geometry
 which describes a conformal theory. A
conformal theory on $S^1\times R^3$ with circumference $\beta$ for $S^1$ has no phase transition since the value of $\beta$ can be scaled away by
conformal invariance \cite{Witten:1998zw} $-$ i.e the vacuum phase is equivalent to the thermal phase. 

Even with $M\neq 0$, when we do not have any D7-branes $-$ i.e
 we do not have any matter in the fundamental representation\footnote{The field theory has  bi-fundamental fields $A_i, B_j$ and in  the far
 IR can be equivalently described by pure glue $SU({M})$ theory. If $T_c$ is very small, the confined phase consists of  glue balls
 and the deconfined phase consists of free gluons of $SU({M})$. If $T_c$ is large, the deconfined phase is best described by $A_i, B_j$
 fields.} $-$ the confinement to deconfinement phase transition
for the gauge theory mimics the first order transition in pure glue theory and is described by a Hawking-Page transition in the dual geometry. 

Observe that in deriving (\ref{actiondiff}), we defined the boundary $r={\cal R}\rightarrow \infty$, but {\it did not} explicitly add a UV geometry. 
By adding counter terms $S_{\rm counter}$ to the on-shell action, we subtracted the terms in $S_{IIB}+S_{GH}$ that diverge at the
boundary $r={\cal R}$, which is effectively choosing a particular UV completion. Explicitly adding an AdS UV cap would require taking account
of the localized sources in the bulk in addition to the fluxes, and the exact on-shell action for a UV complete geometry is not known. 
However, the UV completion resulting from our regularization already gives us a first order phase transition with an exact result for the
critical temperature and thus is already insightful. Furthermore, since confinement is an IR phenomenon, the critical temperature may not be
extremely sensitive to the details of the UV completion and thus the $T_c$ in (\ref{T_c}) can even be relevant for the UV complete geometry.  
    
\section{Conclusion and discussions}

In this paper we have managed to tie up some of the loose ends of our earlier works \cite{FEP, jpsi, vaidya} related to the gauge theory description of the UV complete
geometry predicted in the gravity side. The RG flow from UV to IR at zero temperature shows how the conformal behavior in the far UV ties up with the confining 
dynamics in the far IR. The intermediate-energy
physics is more involved and will be elucidated in our upcoming work \cite{toappear} where we will also discuss how to evaluate 
the spectrum of the theory. As an interesting outcome of the UV completion, we could see how the stability of our background could be justified. Furthermore 
phase transition and related IR issues appear naturally in our set-up. If we ignore the flavor branes, our gravity description gives us a first-order phase transition. 
Further details on this will appear in \cite{fangmia}. In the presence of the flavor branes, the physics is slightly more involved and will be discussed in \cite{toappear}. 
We have also managed to compare our model with some of the other models that study large $N$ thermal QCD 
and showed how certain calculations may become more tractable in our set-up. Whether this is true for most of the other details of large $N$ 
thermal QCD remains to be seen.  

{\bf Acknowledgement}: We would like to thank M. Gyulassy for helpful discussions and especially G. Mandal and T. Morita for patiently explaining their 
recent paper \cite{manmort}. The work of L. C and K. D is supported in part by the NSERC grant, the work of F. C is supported in part by the Schulich grant, 
the work of O. T is supported in part by the FQRNT grant and the work of M. M is supported in part by the Office of the Nuclear Science of the US DOE grant 
number DE-FGO2-93ER40764.

\end{document}